\documentclass[twocolumn,showpacs,preprintnumbers,
amsmath,amssymb,superscriptaddress]{revtex4}

\usepackage{graphicx}

\begin{document}

\title
{Molecular electronics exploiting sharp structure in the
electrode  density-of-states.
 Negative differential resistance and
Resonant Tunneling in a poled molecular layer on
Al/LiF electrodes.
}

\author
{
Z.H. Lu}
\affiliation{Department of Materials Science and Engineering, University of Toronto\\
184 College Street, Toronto, Ontario M5S 3E4
}
\author
{ M.W.C. Dharma-wardana$^*$
}
\affiliation{
National Research Council of Canada, Ottawa,Canada. K1A 0R6\\
}

\author
{ R.S. Khangura}
\affiliation{Department of Materials Science and Engineering, University of Toronto\\
184 College Street, Toronto, Ontario M5S 3E4
}
\author
{Marek Z. Zgierski}
\affiliation{
National Research Council of Canada, Ottawa,Canada. K1A 0R6\\
}
\author
{Douglas Ritchie}
\affiliation{
National Research Council of Canada, Ottawa,Canada. K1A 0R6\\
}
\date{\today}
\begin{abstract}
Density-functional calculations are used to clarify the role of an
ultrathin $LiF$ layer on $Al$ electrodes used in molecular
electronics. The $LiF$ layer creates a sharp density of states (DOS),
as in a scanning-tunneling microscope
(STM) tip.  The sharp DOS,
coupled with the DOS of the molecule leads
 to negative differential resistance (NDR).
Electron transfer between oriented  molecules
 occurs {\it via} 
resonant tunneling. The  $I-V$ characteristic for a
 thin-film of tris (8-hydroxyquinoline)-
aluminum ($AlQ$) molecules, oriented using 
electric-field poling, and sandwiched between two $Al/LiF$ electrodes
is in excellent agreement with theory.
This molecular device presents a new paradigm 
for a convenient, robust, inexpensive alternative to
 STM or mechanical 
break-junction structures.
\end{abstract}
\pacs{PACS Numbers: 05.30.Fk, 71.10.+x, 71.45.Gm}
%
\maketitle
%
{\it Introduction.}
An intense effort has been directed to 
the practical realization of molecular electronics (ME)  
\cite{gimzew,datta,ratner},
since conventional silicon technology is
approaching the limit  of ``Moore's
 law''~\cite{moore}. The advantages of ME are 
 related to properties unique to
 molecules. Molecules are
nanostructures 
with near-perfect topological linkages. The disruption
in surface topology  involved in microfabrication
is a limiting factor in using inorganic
 materials such as GaAs\cite{surfacedis}.
The energy scales associated with molecules (unlike, say, quantum dots)
make them capable of operating at room temperature.
Mechanically controllable break junctions\cite{read} and molecular contacts 
using scanning-tunneling microscope (STM) tips have been 
considered\cite{gimzew}.
However,  molecule-atom contacts
may sometimes take disruptive pathways that lead to fragmentation
 of the metal contacts or the 
 molecule itself\cite{critique,turak}. Assuming non-disruptive
 contacts  could be made, an additional step of coherently
connecting many molecules together 
 requires exotic tools.
Hence the future of ME depends on  surmounting the
theoretical and practical
challenge of realizing robust, reproducible, inexpensive devices.

In this study we show both theoretically and experimentally that
a practical, robust realization of 
molecular-NDR characteristics is possible by exploiting two important
features, {\it viz}., (i) the 
special characteristics of the density of states (DOS) of $LiF$
thin films deposited on $Al$, and
(ii) The possibility of aligning molecules by 
electric poling, i.e,  cooperatively aligning the molecules along the
 field direction when
an external electric field is applied.
 Examples of electric poling are common in the field of
 liquid
crystals and non-linear optical polymers\cite{polling}. 
  The experimental $I-V$ characteristic
inclusive of NDR effects is found to be in agreement  
with our calculations.
%
%
%

{\it The device structure.--}
The experimental structure
 was fabricated  in a cluster tool having a central
distribution chamber, a load-lock chamber, a plasma cleaning chamber, a
sputtering chamber, an organic molecular-growth chamber, and a
metallization chamber.
 Figure 1 shows the layer sequence used.
 A 2"x2" $Si$ substrate with $\sim$ 70 nm of 
oxide was
cleaned by oxygen plasma and then transferred to the 
 metallization chamber
for electrode deposition. The bottom $Al$ electrode (1 mm wide and 2 mm
spacing between adjacent electrodes) was fabricated by thermal
evaporation through a shadow mask in the metallization chamber. The top
electrode  was
 similarly deposited  but the electrode
lines run orthogonal to the bottom ones. The interception area
(1mm x1mm) forms one individual device. One hundred such testing devices
were fabricated on a  2"x2" silicon substrate. After the
deposition of the bottom electrode,
the wafer was transferred to the organic molecular-growth
chamber for $AlQ$ deposition. Electronic grade $AlQ$ was
purchased from Kodak. 
 $AlQ$ films were fabricated 
using a low-temperature K-cell with an alumina crucible. Thicknesses of
50-200 nm have been tried and 100 nm $AlQ$
was optimal with good electrical properties and high 
yield ($>$90\%). The 100 nm $AlQ$ thin-film
devices used $Al$ electrodes having
an ultrathin layer of ( $\sim 0.2$-0.3 nm)  $LiF$ between
 $Al$ and $AlQ$ \cite{grosea}.
 The crucial role of the
$LiF$ will be discussed below.
 After the
deposition of the top electrode, the device was sealed off
from the ambient using a  120 nm 
silicon oxide passivation layer.
This passivation layer is very critical to obtaining 
reproducible results.
\begin {figure}
\includegraphics*[width=7.0cm, height=8.0cm]{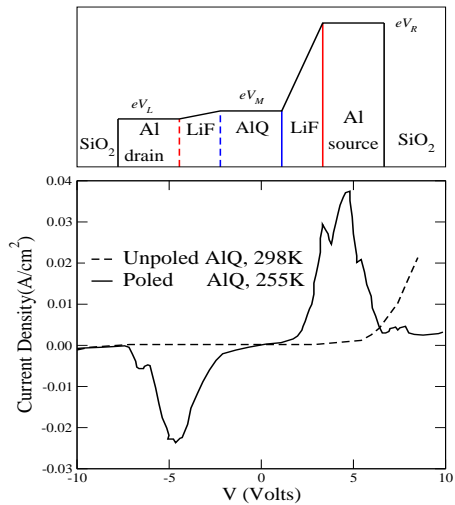}
\caption
{ Top: Schematic of the molecular device, and voltage
profile in the device for poled AlQ. The ''bottom'' Al-electrode
is at the right end.
Lower panel: $I-V$ characteristics of the device without
electric poling and with poling where NDR  structures arise.
}
\label{fz}
\end{figure}

{\it  Measurements and results.--}
The current-voltage ($I-V$) measurements were
conducted using a HP4140B meter with a Materials Development probe
station. Figure 1 (lower panel) shows $I-V$ characteristics of one typical
device before poling (dashed line) at 298 K and after poling (solid
line) at 255 K. The electric poling was made by sweeping the applied
voltage from the starting poling voltage $V_p$  onwards
 with a dwell time of 1 second per voltage point,
in the same mode of $I-V$ measurement. The starting $V_p$ is 
  $\sim 13$ V at 298K
and $\sim 10$ V at 255K. The maximum poling voltage is $\sim 18$ V. It is quite
clear that after electric poling, sharp NDR peaks  near -3 V and +3V were
developed.  The peak structures are very similar to those observed on a
single molecule or self-assembled molecular monolayer\cite{datta,read}.
 We also
found that the NDR peak can be erased by passing a high current at a
higher positive bias-voltage, and the erased peak can be re-generated by
poling. It can be repeated many times on some devices. Because there are
no strong primary bonds between molecules, the molecular alignment may
be readily destroyed by thermal effects. Nevertheless we found that over
90\% of the device showed similar characteristics.
Similar, sporadic and unexplained observations of NDR from AlQ has
been reported in the literature\cite{kim}.
 Here we present controlled
observations whose origin is quantitatively explained.

{\it Theoretical considerations.--}
 The electrode (see Fig. 1) chemical potentials are
  $\mu_L$, and  $\mu_R$, with 
 $-|e|V=\mu_L-\mu_R$, where $V=V_R-V_L$ is the applied potential.
  The
 electron flow  involves (i) electron injection
 into the molecular layer  at the right, denoted by $R\to M$,
and described by a transmission function $T_{RM}$ which is essentially
the joint density of states of the electrode surface and the molecular
layer, weighted by a matrix element.
  (ii) Transfer of carriers from
 molecule to molecule, $M\to M$, described by $T_{MM}$ and finally,
 (iii) transfer from  the
 the molecular layer to the left electrode, $M \to L$, described by $T_{ML}$.

We begin by a qualitative discussion.
 
 The interaction between $Al$ surfaces and organic molecules like
 $AlQ$ has been studied extensively\cite{mason,gennip}. The  $AlQ$
 molecules placed directly in contact with the $Al$ surface is known
 to undergo reactions (e.g.,between $Al$ and the
 quinolate-$O$). Clearly, an important role of the $LiF$ is
 in shielding the molecules against such reactions.
 Various indirect interfaces, e.g, $Al/X/AlQ$, where X may be, 
 say, $LiF$, $CaF_2$, $SiO_2$ etc., have been examined. It has been found
 that a few atomic layers of $LiF$ separating the $AlQ$ from the
 $Al$ electrode produces a striking improvement in the $I-V$
 characteristics and lowers the threshold for
 electron injection. Although this system has been extensively investigated,
 a clear explanation of how  $LiF$ helps has not been forthcoming.

 We show that the $Al/LiF$-electrode DOS is like a $\delta$-function
sitting on the slowly varying DOS of the $Al$ substrate. The DOS of the
$AlQ$ molecule is also found to have a strong sharp feature (SF) in the
unoccupied manifold of states (UMS). At the appropriate bias, electrons
are injected into the  SF of the UMS. Unlike injection into the 
lowest unoccupied state (``LUMO'') which is affected by the Coulomb blockade
and the need for molecular rearrangement, injection to a higher energy
UMS couples only weakly to the ground-state electron distribution.
The injected electrons rapidly migrate in the poled molecular film by
resonant tunneling. Hence the inter-molecular transmission function
$T_{MM}$ is essentially
unity. Finally, electrons arriving at the drain electrode
are in an excited state above the Fermi energy of the $Al$ electrode,
and can resonantly tunnel into the slowly varying 
unoccupied DOS of the $Al$ substrate. This resonant process
also involves a negligible potential drop if the intervening $LiF$  layer
is atomically thin. Thus essentially all the potential drop occurs
at the injection electrode (see Fig.1, top panel) where there is also
an ``empty'' region (not shown in the figure) 
between the $LiF$ surface and the first
$AlQ$ layer. All these lead to an effective tunneling length $s$ which
will appear in the matrix element connecting the electrode states
and the $AlQ$ states.
Thus the potential profile in
 Fig. 1 is rather schematic and does not show the $LiF-AlQ$ tunneling gap
 which is found from our calculations (see below) to be about $\sim 0.1$ nm,
while the $LiF$ layer is about 0.2-0.3 nm.
 We have, to
leading order,
\begin{equation}
T_{RM}(E,V)=(2\pi)^2|A_{RM}(V)|^2\rho_R(E-eV)\rho_M(E)
\end{equation}
Here $A_{RM}$ is the matrix element connecting the molecule
and the electrode, and $\rho_R,\, \rho_M$ are the DOS of the
right electrode (R) and the molecule (M) respectively.
Given our previous conclusion that
$T_{MM}$, and $T_{ML}$ are resonant processes of the
order of unity, the current is effectively determined by
$T_{RM}$. The bias dependence of $A_{RM}(V)$ can be neglected
only if two sharp features in the two DOS functions are involved.
However, the bias dependence of $A_{RM}(V)$ is
important, and this can be included by evaluating the matrix element
as in Lang\cite{lang}.

 A standard Fermi Golden-rule analysis, or a
 Landauer-B\u{u}ttiker approach may be used to write an expression for the
 current $I$ in terms of the transmission function $T(E,V)$.
 \begin{equation}
 I=2\sigma_0\int_{-\infty}^{\infty}dET(E,V)[f(E-\mu_R)-f(E-\mu_L)]\\
  \end{equation}
  Here $\sigma_0$ is the quantum of conductance $e^2/h=1$ 
  in atomic units, which corresponds to a resistance of 12.9 $K\Omega$
  in conventional units.
  The Fermi factors
are effectively step functions even at room temperature.
 Thus the
dominant contribution to the current is given by:
\begin{equation}
\label{i-v}
I(V)=A\int_{\mu_L}^{\mu_R}dE\,\rho_R(E-eV)\rho_M(E)
e^{-2s\{2(E-W)+eV\}^{1/2}}
 \end{equation}
where $A$ is a numerical factor and $\mu_L=E_F$ is the Fermi energy,
while $\mu_R=E_F+eV$.
 Here $s$ is an effective tunneling length
linking the molecule and the $Al/LiF$ electrode, and
$W$ is the workfunction\cite{lang}. These are
treated as parameters of the problem. The main effect of higher temperatures
comes in via a convolution of the $T_{MM}$ which involves 
molecular Debye-Waller effects. We have neglected such effects.

The left electrode was made by depositing $LiF$ and then $Al$
on $AlQ$, and hence follows the reverse of the fabrication sequence
used for the right electrode. However, the electronic processes
in the device are essentially symmetric, with slightly
different values for $s$, $W$ etc. When the bias is
reversed, the carrier processes reverse direction.
This is different to the situation
in $Au-thiol\,$-STM devices where the thiol is chemically attached
to the $Au$ electrode. Then 
NDR is obtained by having two potential drops at the
two electrodes\cite{datta}. Also, unlike in the  $Au-thiol\,$-STM
system, the $AlQ$ molecules are shielded from the $Al$ electrode by
the $LiF$ overlayers, and the main potential drop
occurs at the injector $LiF$ dielectric layer, at the empty region
between the $LiF$ surface and the the first $AlQ$ moleculer layer.
This is  estimated to be about $\sim 1.2$ Angstroms and
plays the role of a tunneling length.

\begin{figure}
\includegraphics*[width=9.0cm, height=10.0cm]{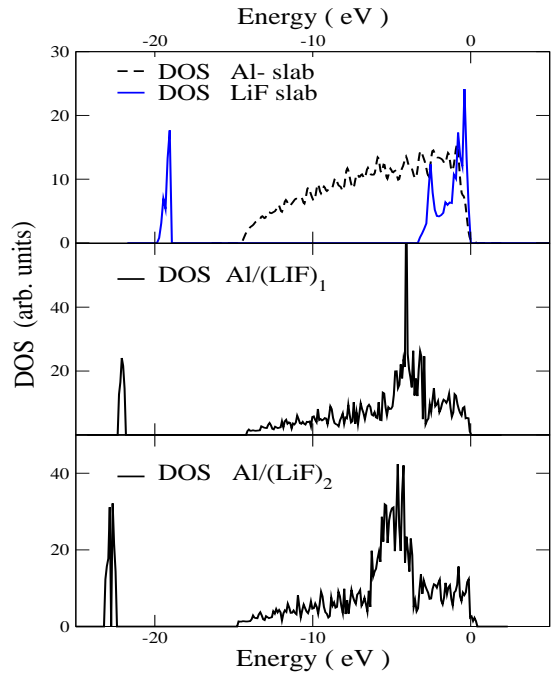}
\caption
{The top panel contains the occupied-state DOS of bulk-Al and bulk-LiF,
with both Fermi levels  set to zero. The lower two panels show the
DOS with  a single  monoatomic layer of LiF, and 2 atomic layers of LiF,
deposited on Al and structure optimized by total-energy minimzation.
}
\label{figdos}
\end{figure}

  {\it Density of states of the Al/LIF electrode.--}
  Bulk $LiF$ and bulk $Al$ both have FCC structures and are very nearly lattice
  matched. While $Al$ is a very good metal, $LiF$ is a large-bandgap insulating
  ionic crystal. We have carried out first-principles density functional
  calculations to simulate $Al/(LiF)_n$ where the number of atomic planes
  $n$ was varied from 1 to 4 to give LiF-overlayers. The crystallographic
  details of the experimental $Al$ surface is unknown. We have carried out
  calculations with the $LiF$ layers arranged along  [001] as well
  as along [111] directions with very
  similar results.
  The $Al$ substrate was modeled  with 6 planer layers of $Al$ in the growth
   direction, and periodically continued in the $x-y$ directions.
  The last layer of $Al$ and the $LiF$ layer were geometry optimized using
 standard
  energy minimization methods available with the VASP plane-wave
  code\cite{vasp}. The simulation cell (SC) included five vacuum layers
  to separate the $Al$ substrate from the $LiF$. 
  DOS calculations were done  using an
  $11\times 11\times 11$ Monkhorst-Pack $k$-sampling scheme.
 The $Al$ atomic layer
  adjacent to the $LiF$ suffers virtually no reconstruction,
   while the $F$ and $Li$ atoms
  which were initially coplaner in the (001) layers,
   reconstruct in opposite directions, with the $F^-$
  moving inwards, towards the $Al$ layer, while the $Li^+$  ions move outwards,
  along the growth axis.
  This reconstruction is very small ($\sim 0.1 A$).
 Similar effects, as well as
  metal-induced surface states, have been reported in $Cu/LiCl$ and
  related systems recently\cite{arita}.
   A crucial effect of the $LiF$
  layer on $Al$ is seen in Fig.2. The DOS of $Al$ by itself,
 $LiF$ by itself, and in the combined structure
  $Al/(LiF)_n$ are shown in Fig.~\ref{figdos}, all referred to the
   same Fermi level
  set to zero for clarity. It is clear that a sharp structure (similar to that
  in the DOS of an STM tip) is produced in the $Al/(LiF)_n$ system
  if $n$ is small. Thus, if an $Al$ electrode has an {\it atomically thin}
  overlayer of $LiF$, we have an injection DOS similar to those
   created using
  mechanical break junctions or STM-tip molecular contacts.
  
  The calculation of the transmission function $T(E,V)$ requires the DOS
  of the $AlQ$ molecule as well. Here a cluster-type calculation is
appropriate, while it would {\it not} be appropriate for the $Al$ metallic
layers.
The electronic structure of the $AlQ$ molecule  and the $AlQ^-$
negative ion were calculated using the Gausssian-98 code at the
B3LYP/3-21G* level\cite{acro,G98}. These calculations included geometry
optimization together with total energy minimization and state of the
art gradient corrected exchange-correlation functionals. Density of states
(DOS) curves were 
 constructed using a 0.25 eV Lorentz broadening of the
one-particle eigenenergies.
 These calculations reveal a highest occupied
molecular orbital (HOMO) at -5.02 eV and a lowest unoccupied molecular
orbital (LUMO) at -1.79 eV, while the optical gap was previously
calculated by us~\cite{marekgap} to be 2.7 eV.

\begin{figure}
\includegraphics*[width=7.0cm, height=8.0cm]{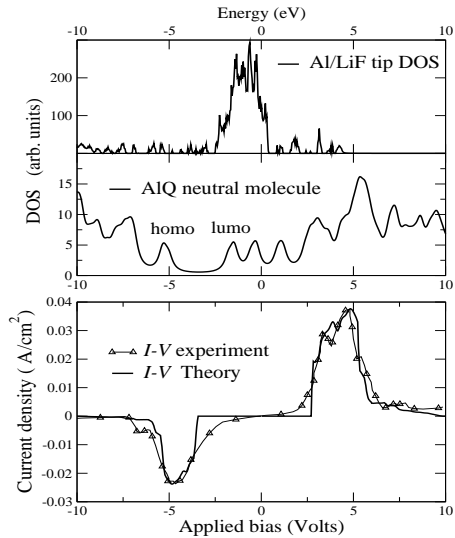}
\caption
{ Top: 
the $Al/LiF$ ''tip'' like DOS with the $Al$ DOS subtracted out.
Middle panel shows the DOS of the neutral $AlQ$ molecule.
Lower panel: $I-V$ characteristics of the device
compared with theory.
}
\label{expth}
\end{figure}

{\it Comparison of calculated and observed $I-V$ characteristics.--}
The analysis leading to Eq.~\ref{i-v} showed that the $I-V$ characteristic
would depend on the main features of the DOS of the molecular film, i.e.,
the molecule itself as we have a simple chemically uncoupled array of
molecules.
In fig.~\ref{expth}, top panel, we present the ``tip-like'' 
 DOS of the $Al/LiF$ layer, 
with the
background $Al$ density of states subtracted off (e.g, bottom panel
 of Fig.2 minus
the $Al$-slab DOS from the top panel of Fig.2).  The DOS of the
$AlQ$ molecule is shown in the middle panel.
 The bottom panel shows
 the experimental $I-V$ data at 255 K
compared with  the  theoretical result obtained from Eq.~\ref{i-v}.
The  experimental $I-V$  agreed with the theory for 
a workfunction  $W\sim 4.6-4.9 eV $,  and
 with the effective tunneling length $s\sim 0.8-1.3$
  Angstroms. 
   The numerical factor $A$ was $\sim 0.001$.

{\it Discussion.--}
In the usual picture ({\it not used by us}$\,$)
of electron transport across the A/AlQ/B structures
under bias, it is assumed that an electron is injected
from the source 
 into the LUMO of the $AlQ$ molecule,
 converting it to a transient $AlQ^{-*}$ anion.
This involves Coulomb blockade, 
rearrangement of bond lengths,
bond angles etc.,  to give the actual  $AlQ^-$ anion. 
 The carrier then hops to a
neighboring $AlQ$ molecule and moves towards the drain electrode.
There is no resonant transfer possible not only because the $AlQ$ 
molecules are not properly oriented, but also because the
available  LUMO states are not in resonance with the $AlQ^-$
eigenstate containing the carrier electron. The $I-V$ characteristics
show no NDR effects.

In contrast, in the structure discussed here,
the  molecules in the electric-field poled $AlQ$ layer form an oriented
array. The electron is injected to an unoccupied high-energy state
 by the sharp DOS structure of the $Al/LiF$ electrode.
The electron reaches the drain electrode by inter-molecular resonant transfer.
Since only high-energy unoccupied states are invoked, this process 
does not involve significant Coulomb blockade or
bond-rearrangement bottlenecks.
The device is robust, easy to make and works at ambient
temperatures.

{\it Conclusion--} We have shown, experimentally and theoretically, that
a robust room-temperature molecular device, showing $I-V$ characteristics
similar to those obtained in STM-based molecular devices,
can be fabricated using $Al/LlF$ electrodes and a poled molecular film.
New insight into the role of $LiF$ overlayers in electrode processes is
also presented.

\end{document}